\documentstyle[12pt]{article}
\renewcommand
\baselinestretch{1.4}

\begin{document}

\hfill{UM-P-94/102}

\hfill{RCHEP-94/27}

\begin{center}
{\LARGE \bf Sphalerons in the Standard Model with a real Higgs singlet}

\vspace{4mm}

\renewcommand
\baselinestretch{0.8}\vspace{4mm}

{\sc J. Choi}\\
\renewcommand
\baselinestretch{1.4}

{\it
Research Centre for High Energy Physics,\\
School of Physics, University of Melbourne, \\
Parkville, Victoria 3052, Australia}
\renewcommand
\baselinestretch{1.4}

\vspace{5mm}
\end{center}

\begin{abstract}
Sphaleron energies within the standard model with a real Higgs singlet
added on are calculated. The coupled non-linear equations of motion
are numerically solved and the sphaleron energy evaluated for a set of
parameters in the Higgs potential. I find a small difference in the
sphaleron energy compared to the standard model. A slightly stronger
constraint on the strength of  the first order phase
transition thus results for this model.
\end{abstract}

\section{Introduction}
If the baryon asymmetry of the universe was generated at the electroweak
phase transition, as many models are now speculating \cite{nels}, it is
important that the rate of baryon violation be determined more accurately.
This rate in the broken symmetry phase is directly related to the energy
of the sphaleron configuration in the model, because the sphaleron energy
sets the height of the energy barrier between the topologically inequivalent
vacua, and baryon violation occurs whenever such a vacuum transition takes
place. To preserve any baryon asymmetry created at the phase transition
the baryon violation rate must be suppressed in the broken phase and this
imposes a constraint on the sphaleron energy at the phase transition. Hence
it is important to calculate the sphaleron energy within each model being
considered. It is in any case interesting to know the explicit form of
the sphaleron configuration in extensions of the Standard
Model.

Manton {\it et.al} calculated the sphaleron energy within the Standard Model
(SM) using a radially symmetric ansatz for the gauge and Higgs fields
\cite{mant}. They obtained
\begin{equation}
E_{\rm sph}=\frac{4\pi v}{g}B(\lambda/g^2)
\label{SMsphE}\end{equation}
where $v$ is the Higgs vacuum expectation value (VEV), $g$ is the weak
gauge coupling and $\lambda$ is the quartic self coupling of the Higgs
boson. Their calculations showed $1.5\leq B\leq 2.7$ for $\lambda/g^2$
between $0$ and $\infty$.
Since the baryon violation rate in the broken phase is proportional to
$e^{-E_{\rm sph}/T}$, suppression of the baryon violation rate in the
broken phase requires \cite{shap}
\begin{equation}
\frac{E(T_c)}{T_c}\stackrel{>}{\sim} 45,
\label{Econdn}\end{equation}
independent of the model, where $T_c$ is the critical
temperature for the phase transition. Hence in the SM, one obtains
\begin{equation}
\frac{v}{T_c}\stackrel{>}{\sim} 1.4
\end{equation}
as the condition on the strength of the first order phase transition
required.

Because this constraint seems so restrictive in the SM [apart from other
possible inadequacies like the strength of Kobayashi-Maskawa (KM) CP
violation] it seems likely that an extension of the SM is required to
generate sufficient baryon asymmetry at the electroweak phase transition.
Hence it is important that the sphaleron energy be calculated in these
models. In this paper I consider the sphaleron solutions in the minimal
SM augumented by a real singlet Higgs field. This model was shown to be
capable of satisfying at least one of the several conditions needed for
electroweak baryogenesis, namely the condition that any baryon asymmetry
created at a first order phase transition does not get washed away by
subsequent sphaleron processes \cite{us}. This model is interesting due to
the existence of tree-level trilinear Higgs couplings which seem to enhance
the possibility of satisfying this condition (I also point out other
possible interest this model has in this regard in the discussions later).
Kastening and Zhang
\cite{kast} have investigated sphaleron solutions when a complex Higgs
singlet is added to the SM (their paper also applies to the real Higgs
singlet case but with the trilinear coupling terms missing) and found
non-zero differences in $E_{\rm sph}$ compared to the SM case. I too
want to compare the sphaleron energy of this model with the SM.
If the calculations here show a marked increase in the sphaleron
energy compared to the SM value then one can expect a greater
suppression of the sphaleron rate in the broken phase, relaxing the
constraints on the Higgs masses somewhat. If on the other hand a
considerable decrease in the sphaleron energy is seen in this model, then
the first order phase transition will need to be even stronger and
consequently less parameter space will be available to prevent a washout
of the baryon asymmetry of the universe. The present work is thus a refinement
of Ref.\cite{us}, since the minimal SM sphaleron configuration was used there
as an approximation.

\section{The equations of motion}
A real singlet Higgs field $S$ is added to the minimal SM with its doublet
Higgs field $\phi$. The Lagrangian of the gauge and the Higgs sector of
the model is then:
\begin{equation}
{\cal L}=-\frac{1}{4}F^{\mu\nu}_a F^{\mu\nu}_a +
         (D_{\mu}\phi)^{\dagger}(D^{\mu}\phi) +
         \frac{1}{2}\partial_{\mu}S\partial^{\mu}S - V(\phi,S),
\end{equation}
where
\begin{equation}
F^{\mu\nu}_a = \partial^{\mu}W^{\nu}_a - \partial^{\nu}W^{\mu}_a +
               g\epsilon_{abc}W^{\mu}_b W^{\nu}_c
\end{equation}
\begin{equation}
V(\phi, S) = \lambda_{\phi}(\phi^{\dagger}\phi)^2
                -\mu_{\phi}^2\phi^{\dagger}\phi
                + \frac{\lambda_S}{2}S^4 - \frac{\mu_S^2}{2}S^2
                -\frac{\alpha}{3}S^3 + 2\lambda(\phi^{\dagger}\phi)S^2
                - \frac{\sigma}{2}(\phi^{\dagger}\phi)S.
\label{V}\end{equation}
\begin{equation}
D_{\mu}\phi = (\partial_{\mu}-\frac{1}{2}ig\tau^a W_{\mu}^a)\phi.
\end{equation}
(For simplicity, the fermion fields will be set to zero as well as the
${\rm U_Y(1)}$ gauge field by setting the weak mixing angle to zero.)

The Higgs potential $V(\phi,S)$ breaks the ${\rm SU_L(2)\otimes U_Y(1)}$
symmetry to ${\rm U_Q(1)}$ when the Higgs fields take on the VEV's at:
\begin{equation}
\langle\phi(x)\rangle = \frac{1}{\sqrt{2}}
 \left(\begin{array}{c}0\\u \end{array}\right),
 \quad\mbox{and}\quad \langle S(x)\rangle = \frac{1}{\sqrt{2}}v,
\end{equation}
where $u = 246$ GeV and $v$ is kept non-zero for full generality.
The minimum of the potential (\ref{V}) is given by $(u,v)$
where $u,v$ are solutions to
\begin{equation}\begin{array}{l}
(-\mu_{\phi}^2 + \lambda_{\phi}u^2 + \lambda v^2 - \sigma v/2)u=0\\
-\mu_S^2 v + \lambda_S v^3 + \lambda v u^2 - \sigma u^2/4 - \alpha v^2=0.
\end{array}\label{mineq}\end{equation}
I eliminate the parameters $\mu_{\phi}^2$ and $\mu_S^2$ in favour of
one of the $(u,v)$ solutions to Eq.(\ref{mineq}):
\begin{equation}\begin{array}{l}
\mu_{\phi}^2 = \lambda_{\phi}u^2 + \lambda v^2 - \sigma v/2 \\
\mu_S^2 = \lambda_S v^2 + \lambda u^2 - \sigma u^2/4v - \alpha v.
\end{array}\label{mueq}\end{equation}

Now for static classical fields, one has
\begin{equation}
\partial_0 F^{\mu\nu}_a = \partial_0\phi =  \partial_0 S = 0 \qquad
\mbox{and}\qquad W^a_0 = 0.
\end{equation}
Then the equations of motion derived from ${\cal L}$ are:
\begin{eqnarray}
(D_j F_{ij})^a & = & - \frac{i}{2}g[\phi^{\dagger}\tau^a(D_i\phi)-
                  (D_i\phi)^{\dagger}\tau^a\phi]
\\
D_i D_i \phi & = & [2\lambda_{\phi}(\phi^{\dagger}\phi) - \mu_{\phi}^2]\phi
               + 2\lambda S^2\phi - \frac{\sigma}{2}S\phi
\\
\partial_i \partial_i S & = & [8\lambda\phi^{\dagger}\phi - 2\mu_S^2]S
	+ 4\lambda_S S^3 - \sigma\phi^{\dagger}\phi - 2\alpha S^2.
\end{eqnarray}
And the corresponding energy functional is:
\begin{equation}
E = \int d^3x [\frac{1}{4}F^a_{ij}F^a_{ij} + (D_i\phi)^{\dagger}(D_i\phi)
    + \frac{1}{2}(\partial_i S)(\partial_i S) + V(\phi,S) ].
\end{equation}

Now I use the spherically symmetric ansatz of the form:
\begin{eqnarray}
W_i^a\tau^a dx^i & = & -\frac{2i}{g}f(gVr)dU^{\infty}(U^{\infty})^{-1}
\\
\phi & = & \frac{u}{\sqrt{2}}h(gVr)U^{\infty}
       \left(\begin{array}{c}0\\1\end{array}\right)
\\
S & = & \frac{v}{\sqrt{2}}p(gVr),
\end{eqnarray}
where $\qquad U^{\infty} = \frac{1}{r}\left(\begin{array}{cc} z & x+iy \\
-x+iy & z \end{array}\right)\qquad$ and $\qquad V = \sqrt{u^2+v^2}.$

Then the equations of motion become:
\begin{eqnarray}
\xi^2\frac{d^2f}{d\xi^2} & = & 2f(1-f)(1-2f)
	- \frac{\xi^2 u^2}{4v^2}h^2(1-f)
\nonumber\\
\frac{d}{d\xi}\left(\xi^2\frac{dh}{d\xi}\right) & = & 2h(1-f)^2 +
  \frac{\xi^2}{g^2 V^2}\left(\lambda_{\phi}u^2 h^3 - \mu_{\phi}^2 h +
  \lambda v^2 p^2 h - \frac{\sigma v}{2\sqrt{2}}ph\right)\label{eqns}
\\
\frac{d}{d\xi}\left(\xi^2\frac{dp}{d\xi}\right) & = & \frac{4\xi^2}{g^2 V^2}
  \left[(\lambda u^2 h^2 - \mu_S^2/2)p + \frac{\lambda_S}{2}v^2 p^3
  	- \frac{\alpha v}{2\sqrt{2}}p^2
	- \frac{\sigma u^2}{4\sqrt{2}v}h^2 \right],
\nonumber\end{eqnarray}
where $\xi\equiv gVr$.
And the energy functional becomes
\begin{equation}\begin{array}{ll}
E = \frac{4\pi V}{g}\int_0^{\infty}d\xi\left\{ \right.&
    4\left(\frac{df}{d\xi}\right)^2 + \frac{8}{\xi^2}[f(1-f)]^2 +
    \frac{\xi^2 u^2}{2V^2}\left(\frac{dh}{d\xi}\right)^2 +
    \frac{u^2}{V^2}[h(1-f)]^2 \\  & +
    \frac{\xi^2 v^2}{4V^2}\left(\frac{dp}{d\xi}\right)^2 +
    \frac{\xi^2}{4g^2 V^4}\left[\lambda_{\phi}(u^2 h^2 -
     \mu_{\phi}^2/\lambda_{\phi})^2 +
	(2\lambda u^2 h^2 - \mu_S^2)v^2 p^2\right. \\
	& + \left. \left.
    \frac{\lambda_S v^4}{2}p^4 - \frac{\sigma u^2 v}{\sqrt{2}}h^2 p
    - \frac{\alpha\sqrt{2}v^3}{3}p^3\right]-\epsilon_o\right\},
\end{array}\label{E}\end{equation}
where $\epsilon_o$ is the normalisation needed to set the minimum value
of $V(\phi,S)$ at zero.

Hence within the validity of the ansatz being used, the problem of
finding the sphaleron energy in this model becomes equivalent to
either minimizing the above energy functional (\ref{E}) or solving the
coupled non-linear differential equations (\ref{eqns}).
The latter method will be used in this paper, by numerically solving
the equations and then calculating the value of $E$ which corresponds
to the solutions found.

Boundary conditions for the functions are required and to obtain
these at $\xi=0$, the equations are Taylor expanded about $\xi=0$ and
regularity is imposed there to obtain
\begin{equation}
\xi\rightarrow 0: \qquad f\rightarrow a_s \xi^2,\quad h\rightarrow
 b_s\xi,\quad p\rightarrow c_s + c_2\frac{\xi^2}{2},
\end{equation}
where
\begin{equation}
c_2 = -\frac{c_s}{3g^2 v V^2}\left[-\frac{\sigma}{\sqrt{2}}u^2
     + 4\lambda u^2 v - \sqrt{2}\alpha v^2(1-c_s)
     + 2\lambda_S v^3(1-c_s^2)\right],
\end{equation}
while at $\xi=\infty$ the Higgs fields are made to take on their zero
temperature VEV's by $f,h,p\rightarrow 1$. The equations are actually
singular at both boundaries, so for large $\xi$ one looks for asymptotic
forms of the functions. I find
\begin{equation}
\xi\rightarrow\infty: \qquad f\rightarrow 1-a_l e^{-a\xi},\quad
 h\rightarrow 1-b_l e^{-b\xi}/\xi,\quad p\rightarrow 1-c_l e^{-c\xi}/\xi,
\end{equation}
where
\begin{equation}
a = \sqrt{\frac{u^2}{4V^2}},\quad
b = \sqrt{\frac{2\lambda_{\phi} u^2}{g^2 V^2}},\quad
c = \sqrt{\frac{4}{g^2 V^2}\left[-\frac{\alpha v}{\sqrt{2}}
     + 2\lambda_S v^2 + \frac{\sigma u^2}{2\sqrt{2}v}\right]}.
\end{equation}

There are three second order ODEs, so one needs to have six boundary
conditions. The variables $(a_s,b_s,c_s,a_l,b_l,c_l)$ implicitly contain
these boundary conditions so the solutions to the equations of motion
and these variables need to be determined simultaneously.

\section{Solutions and results}
Because there is no analytical way of finding the solutions to the
coupled non-linear ODEs of (\ref{eqns}), numerical methods must be used.
In order to do this, some values of the parameters in the potential need
to be chosen, and I choose the same set here as was used in Ref.\cite{us}
for easy comparison. These sets of parameters then indicate
the values for which the first order phase transition is strong enough
while the mass of the smallest Higgs is above the experimental lower
limit, where the minimal SM sphaleron energy is used as an approximation.
Here I improve on this by using a corrected sphaleron energy.

I used the shooting method to solve the equations of motion. The
results obtained for each set of parameters are tabulated in Table 1,
and a typical solution is shown in Fig.1.
Note that the values of $E$ are given here in units of $4\pi V/g$ where
$V=\sqrt{u^2+v^2}$, unlike the SM values which are measured in units
of $4\pi u/g$. The results show that the sphaleron energies in this
model are somewhat lower than the SM case as calculated by Manton
{\it et.al} \cite{mant}. Similarly when compared to the SM + a complex
Higgs singlet case considered by Kastening and Zhang \cite{kast}, I
find lower values of the sphaleron energy.

What does this mean for electroweak baryogenesis? Satisfying
Eq.(\ref{Econdn}) in this model requires
\begin{equation}
\frac{V}{T_c}\stackrel{>}{\sim} \frac{45 g}{4\pi B},
\label{VEVcondn}\end{equation}
where $B$ is now a function of the parameters in this model.
Table II compares the values of $45g/4\pi B$ obtained here with the
$V/T_c$ values obtained in Ref \cite{us} at the phase transition, for
each corresponding set of parameters. They show that the condition
(\ref{VEVcondn}) is easily still satisfied, because the first order
phase transition is still too strong compared to the decrease in the
energy of the sphaleron found.

Of course one must keep in mind all the uncertainties involved in such
an analysis. The rate of sphaleron transitions in the broken phase
involves prefactors to the Boltzmann factor \cite{boch} which is
difficult to calculate near the transition temperature due to infrared
divergences. I have also extrapolated my zero temperature sphaleron
energy up to a high temperature in making the comparison (\ref{VEVcondn})
\cite{brai} and have not taken any quantum corrections into account
\cite{boch}.
Also, I have not considered any source of CP violation in this model at
all. There is still some debate on the question of whether or not the SM
can provide sufficiently strong CP violation in order to produce the
observed baryon asymmetry \cite{shap2}. If it turns out that it can't,
this model being considered here will need a new source of CP violation
\cite{McD}.
If however it turns out that the SM {\it can} provide sufficiently strong
CP violation, then adding the real Higgs singlet will be an extremely
economical way of achieving electroweak baryogenesis as it can provide a
first order phase transition for a larger region of parameter space than
can the minimal SM (Once again there is much uncertainty in the calculations
of the effective potential at finite temperature so hasty conclusions cannot
be made.) It also does not suffer from the possible problem faced by the two
Higgs doublet model which may not be able to produce sufficient baryon
asymmetry in the first place even though it can prevent a washout of any baryon
asymmetry that has been created \cite{hamm}. This is because there are two
VEV's and the phase transition can be in two stages. If after acquiring the
first VEV the sphaleron rate is too suppressed because the gauge symmetry is
broken, not enough baryon asymmetry will be created before the second phase
transition which will preserve this asymmetry. Even though the real Higgs
singlet model considered in this paper also has a two stage phase
transition, because the gauge symmetry remains unbroken after the singlet
acquires its VEV, the baryon violating rate will not be suppressed in
between the two phases.

\section{Conclusion}
I calculated the sphaleron energies in the standard model with a real
Higgs singlet added on, for some representative sets of parameters. These
show a decrease in the actual sphaleron energies compared to the values
found in the Standard Model. However the decrease is not sufficient to
constrain the strength of the possible first order phase transition in
this model. If KM CP violation in the Standard Model is sufficiently strong
for electroweak baryogenesis, then this model will be an extremely economical
way of creating the baryon asymmetry of the universe at the electroweak
phase transition.

\section*{Acknowledgments}
I would like to thank R. Volkas for ideas and numerous discussions. I would
also like to thank the following people for useful hints and
discussions: B. Kastening, J. Gunning, C. Dettmann and K. Liffman. I would
like to acknowledge the support of the Australian Postgraduate
Research Program.

\begin{table}
\caption{Representative parameter values and the corresponding sphaleron
energies (in units of $4\pi V/g$). The parameters $\lambda_{\phi,S}$ and
$\lambda$ are dimensionless while the other parameters have dimension GeV.}
\[\begin{array}{ccccccc}
  \lambda_{\phi}\quad & \lambda_S\quad & \lambda\quad & v\quad
  & \sigma\quad & \alpha\quad & E_{\rm sph}\quad \\ \hline
0.071\quad & 0.082\quad & -0.0234\quad & -125\quad & -12.81\quad &
            12.20\quad & 1.40\\
0.063\quad & 0.073\quad & 0.0250\quad & -154\quad & -13.26\quad &
            3.58\quad & 2.20\\
0.087\quad & 0.090\quad & 0.0114\quad & -245\quad & -9.37\quad &
            1.24\quad & 0.82\\
0.096\quad & 0.082\quad & 0.0499\quad & -146\quad & -17.27\quad &
            5.10\quad & 1.80\\
0.050\quad & 0.077\quad & 0.0168\quad & -117\quad & -12.43\quad &
            8.34\quad & 1.91\\
0.057\quad & 0.0920\quad & 0.0026\quad & -171\quad & -16.39\quad &
            4.64\quad & 0.55\\
0.057\quad & 0.0920\quad & 0.0026\quad & -107\quad & -16.54\quad &
            16.76\quad & 0.91
\end{array}\]
\end{table}

\begin{table}
\caption{Comparison of $V/T_c$ obtained from constraint (2) and
$E_{\rm sph}$ found with $V/T_c$ obtained at the first order electroweak
phase transition.}
\[\begin{array}{ccc}
  45g/4\pi B & m_{\phi,S}\quad & V_c/T_c\quad \\ \hline
1.61\quad & 61,102\quad & 170/107\\
1.03\quad & 73,88\quad & 390/93\\
2.75\quad & 103,108\quad & 510/117\\
1.25\quad & 75,109\quad & 430/97\\
1.18\quad & 67,79\quad & 360/70\\
4.10\quad & 73,95\quad & 360/85\\
2.48\quad & 68,92\quad & 360/63
\end{array}\]
\end{table}

\section*{Figure captions}
\begin{itemize}
\item[Fig.1:]
A typical solution to the coupled non-linear ODE's of Eq.(19).
(The fourth line of parameters in Table 1 was used here.)
\end{itemize}
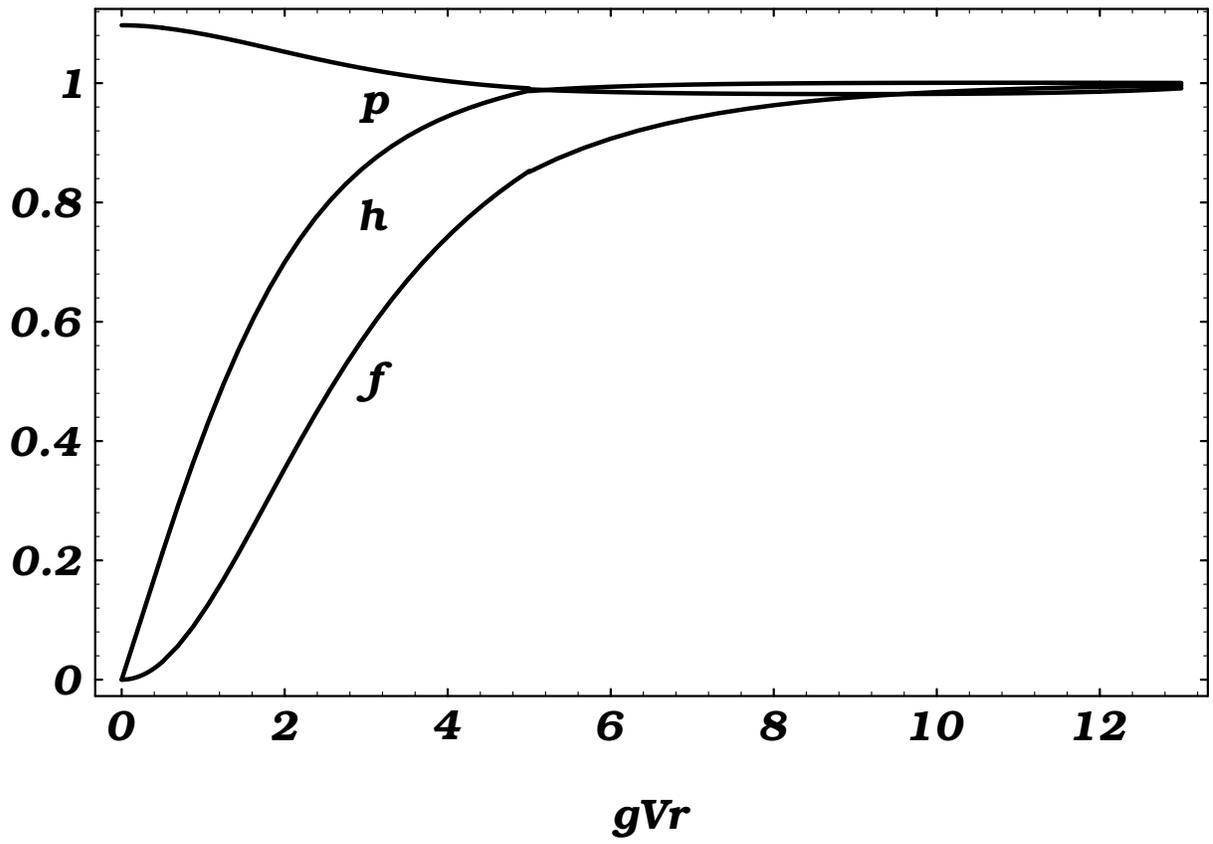
\end{document}